\documentclass[11pt]{article}
\usepackage[usenames,dvipsnames]{color}
\usepackage{hyperref}
\usepackage{amsmath,latexsym}

\begin{document}
\title
{Rotating Killing horizons in generic $F(R)$ gravity theories}
\author
{Sourav Bhattacharya{\footnote{sbhatta@iucaa.in}}\\
Inter-University Centre for Astronomy and Astrophysics (IUCAA)\\ Pune University Campus, Pune - 411007, India}

\maketitle
\abstract
\noindent 
We discuss various properties of  rotating Killing horizons in generic  $F(R)$ theories of gravity in dimension four for spacetimes endowed with two commuting Killing vector fields. Assuming there is no curvature singularity anywhere on or outside the horizon, we construct a suitable $(3+1)$-foliation. We show that similar 
to Einstein's gravity, we must have $T_{ab}k^ak^b=0$ on the Killing horizon, where $k^a$ is a null geodesic tangent to the horizon. For axisymmetric spacetimes, the effective gravitational coupling $\sim\,F'^{-1}(R)$ should usually depend upon the polar coordinate and hence need not necessarily be a constant on the Killing horizon.
We prove that the surface gravity of such a Killing horizon must be a constant, irrespective of whether $F'(R)$ is a constant there or not. We next apply these  results to investigate some further basic features.
In particular, we show that any hairy solution for the real massive vector field in such theories is clearly ruled out, as long as the potential of the scalar field generated in the corresponding Einstein's frame is a positive definite quantity.

\vskip .2cm

{\bf Keywords:} {$F(R)$ gravity, rotating black holes, Killing horizons
\vskip .5cm
  
\maketitle
\section{Introduction}
So far Einstein's theory of gravitation has established its most overwhelming compatibility with observation, 
starting from the solar system to the redshift of type Ia Supernovae, the cosmic microwave background, the galaxy clustering and most recently, the gravity waves~\cite{Abbott:2016blz}. However, the lack of any observational evidence of dark matter particle candidate or the so far elusive nature of the dark energy~\cite{Martin:2012bt, polyakov, ait, woodard} has, in recent times intensified the interest in deviating from the General Relativity at large scales. Such theories are known as the alternative or modified theories of gravity, see~\cite{review, Capozziello:2011et} for recent reviews and exhaustive list of references in this direction.  
   
For theories dealing with alternatives of the dark energy, the goal is
to generate the accelerated cosmic expansion without invoking a cosmological constant by hand. For example, in models like the galileon, a certain scalar field's energy-momentum tensor plays the same role, e.g.~\cite{Martin}. On the other hand, it seems reasonable to hope that such accelerated expansion could  also be generated by merely replacing the Einstein-Hilbert Lagrangian density, $1/(16\pi G)\sqrt{-g}\,\left(R-2\Lambda\right)$, with
$1/(16\pi G)\sqrt{-g}\,F(R)$, where $F(R)$ is some function of the Ricci scalar $R$. Such theories are popularly known as the $F(R)$ theory of gravity which will concern us in this paper.			   

Note however, that generating dark energy-like effects by a given alternative theory and its compatibility with observation does not solve the so called fine tuning problem with the cosmological constant at the quantum level (see e.g.~\cite{polyakov} and references therein). Nevertheless, this does not rule out the interest of such viable alternative models as far as they can fit the observed data and can as well predict something qualitatively new and verifiable. Moreover, it could be argued that such a viable alternative could represent an effective classical theory of a hitherto unknown complete quantum theory of gravitation.

The $F(R)$ theory of gravitation was first introduced in~\cite{Starobinsky} in order to model inflation in the early universe without requiring any strong energy condition violating matter field, via a $R^2$ term in the action. See also~\cite{Cembranos:2008gj} for a demonstration of generation  of non-Baryonic dark matter in $R^2$ gravity theory.  See~\cite{Berry:2011pb} for discussions on gravitational radiation and solar system constraints of linearized $F(R)$ gravity. A review on the inflationary cosmology with the Starobinsky model and its further generalization can be found in~\cite{Bamba:2015uma}.
Aspects of spherical gravitational collapse relevant to this theory can be seen in~\cite{Guo:2013dha, Guo:2015eho, Chakrabarti:2014bsa}.  The Buchdahl limit~\cite{Wald:1984rg} for spherical stars in this theory was discussed in~\cite{Goswami:2015dma}, showing there can be more mass packed inside a star of a given radius, compared to the General Relativity. No go theorem for stars with polytropic equation of state can be found in~\cite{Barausse:2007pn, Mana:2015vqa}
in such theories. We further refer our reader to~\cite{Capozziello}-\cite{Nojiri} and references therein for various directions and reviews in $F(R)$ theory.

In this work we are concerned with black hole spacetimes in generic $F(R)$ theory of gravity (see e.g.~\cite{Canate:2015dda}
and references therein). Static black hole solutions with asymptotically flat boundary condition for different ansatz for the
function $F(R)$, can be seen in~\cite{bh1}-\cite{Aghamohammadi:2010yq}. We further refer our reader to~\cite{Nojiri:2013su, Nojiri:2014jqa} for a discussion on the anti-evaporation of the Schwarzschild-de Sitter solution admitted in the $F(R)$ gravity.

While it is well established in vacuum/Electrovacuum Einstein's theory that the Kerr-Newman family is the unique asymptotically flat black hole solution, e.g.~\cite{Chrusciel:2012jk}, it is not clear so far whether this is also the case for a given $F(R)$ theory as well. We refer our reader to~\cite{bth1}-\cite{bth5} for steps towards understanding this issue for static and spherically symmetric black hole spacetimes. 
A feature related to the black hole  uniqueness properties is the celebrated no hair theorems, stating that there can be no matter field
other than the long range gauge fields at the exterior of a stationary black hole spacetime, so far which has been fairly well understood in the context of the Einstein gravity~\cite{nh1}-\cite{Sudarsky}. We also refer our reader to~\cite{Herdeiro} and references therein for a recent review on this and also its possible violations for complex matter fields
admitting stationary energy-momentum tensor.

We shall focus on rotating spacetimes in generic $F(R)$ theories of gravity in this work. We refer our reader to~\cite{Larranaga:2011fv, Sheykhi:2013oya, Ghosh:2012ji, Sheykhi:2013yga} for exact rotating 
black hole solutions and their thermodynamics with constant or varying with radial distance Ricci scalar. However, note that unlike Einstein's gravity, in this case, in a generic stationary axisymmetric spacetime the Ricci scalar should also depend upon the polar angle. To the best of our knowledge, there exists no work so far addressing
properties of such more general stationary axisymmetric spacetimes.  
In this paper, we shall first discuss some basic geometric properties of stationary axisymmetric Killing horizons admitted by a {\it generic $F(R)$ gravity}. In particular, we shall 
prove that the surface gravity on any regular Killing horizon in such theory must be a constant, irrespective of the fact whether the Ricci scalar is a constant over the horizon or not. Next, using these geometric formalism, we will discuss no hair theorems for the real scalar and massive vector fields.

 Recently, the no hair properties of the scalar field generated in the corresponding Einstein's frame was discussed in~\cite{Canate:2015dda}. It naturally turns out, based on the discussion of the Einstein-Higgs theory~\cite{Sudarsky} that as long as the scalar's potential in the Einstein frame is positive definite, the black hole in the Einstein frame will have no scalar hair. However, in the following by `no hair' and `scalar field' we respectively would  mean no hair for black holes in the $F(R)$ frame, and some additional scalar field, other than the one generated in the Einstein frame by the conformal transformation. However, we shall use the correspondence between these two frames, in particular for the Proca field, for calculational convenience only.

The rest of this paper is organized as follows. In the next section we discuss the field equations and the basic assumptions. Sec.~3 is devoted to establish the properties of the stationary axisymmetric Killing horizons and thermodynamics. Next we discuss the no hair theorems with asymptotically flat boundary conditions. Finally we conclude in Sec.~5.

We work here with mostly positive signature of the metric in dimension four and will set $c=1$ throughout. Our notation for symmetrization or antisymmetrization would respectively be : $T_{(ab)}=T_{ab}+T_{ba}$ and $T_{[ab]}=T_{ab}-T_{ba}$, and so on for higher rank fields.
    
\section{The theory and the basic geometry}
\subsection{The model and field equations}
The action for the $F(R)$ theory of gravity is given by, 
\begin{eqnarray}
S=\int d^4 x \sqrt{-g}\left[ \frac{F(R)}{16 \pi G}+{\cal L}(g_{ab}, \Phi_M) \right],
\label{f1}
\end{eqnarray}
where $F(R)$ is a smooth but otherwise arbitrary function of the Ricci scalar $R$ and ${\cal L}$ collectively stands for
the Lagrangian density for all matter fields $\Phi_M$. The equation of motion corresponding to the metric $g_{ab}$ is given by,
\begin{eqnarray}
F'(R)R_{ab}-\frac12 F(R) g_{ab}-\left(\nabla_a\nabla_b - g_{ab} \Box\right)F'(R)=8\pi G T_{ab},
\label{f2}
\end{eqnarray}
where the `prime' denotes differentiation once with respect to the Ricci scalar, $R$. Setting $F(R)=R$ recovers the General Relativity. Taking the trace of the above equation yields
\begin{eqnarray}
\Box R=\frac{1}{3F''}\left[8\pi G \,T - 3 F''' \left(\nabla_aR\right)\left(\nabla^a R\right)+2F-RF'\right].
\label{f3}
\end{eqnarray}
This equation explicitly shows that even when $T=0$, in an axisymmetric spacetime $R$ would generally be a function of the radial and  polar coordinates, apart from the trivial solution $R=0$.
The above equations of motion are equivalent to that of the Brans-Dicke theory in the Jordan frame with vanishing kinetic term along with a potential $V$ (e.g.~\cite{Canate:2015dda}, and references therein)
\begin{eqnarray}
S=\frac{1}{16\pi G}\int d^4 x \sqrt{-g} \left[\phi R -V(\phi)+ {\cal L}(g_{ab}, \Phi_M)\right], 
\label{f4}
\end{eqnarray}
where $\phi$ is the Brans-Dicke field, $V(\phi)=\phi R(\phi)-F(R(\phi))$ and $ \phi=F'(R)$. We can now go to the Einstein frame by using the conformal transformations, $\Phi=\sqrt{3/16\pi G}\, \ln \phi$ and $\widetilde{g}_{ab}= \exp(\sqrt{16\pi G/3}\,\Phi) g_{ab} $ and throwing away one total divergence to get 
\begin{eqnarray}
S=\int d^4 x \sqrt{-\widetilde{g}}\left[\frac{\widetilde{R}}{16\pi G}-\frac12 \left(\widetilde{\nabla}_a \Phi\right) \left(\widetilde{\nabla}^a \Phi\right) - U(\Phi) + {\cal L}(\widetilde{g}_{ab}, \Phi_M)  \right]. 
\label{f5}
\end{eqnarray}
We shall mostly work in the usual frame~(\ref{f1}). We shall use the above correspondence while dealing with the Proca field only, for mere {\it calculational convenience}. We note that the correspondence between the frames~(\ref{f1}) and~(\ref{f5}) seems to be possible only with $F''(R)>0$~\cite{Canate:2015dda}. We further refer our reader to~\cite{Briscese:2006xu} for an explicit discussion on $F(R)$ models not allowing such correspondence between the two frames. 

We assume that the above conformal transformation is smooth and nowhere vanishing. This means that the causal and horizon structures remain qualitatively unchanged by this. Also, we assume that the asymptotic conditions remain unaffected by such transformation. For asymptotically flat boundary conditions, this means that for both $\widetilde{g}_{ab}$ and $g_{ab}$, the leading behaviour is ${\cal O}(1/r)$, as $r\to \infty$.


We shall not assume any specific form of $F(R)$, but will assume that it is such that it gives rise to regular black hole solutions  with asymptotically flat boundary conditions.

\subsection{The geometry}
The geometric assumptions and methods for the basic set up will be quite model independent and hence similar to the General Relativity, see e.g.~\cite{Wald:1984rg},~\cite{Bhattacharya:2010vr, Bhattacharya:2011dq, Bhattacharya:2013caa} and references therein for details.
We assume that the spacetime is  smooth (i.e., free of any curvature singularity, at least in our region of interest), torsion-free and is stationary and axisymmetric. We refer our reader to~\cite{Elizalde:2010ts} for a discussion on $F(R)$ models free of singularities in the cosmological context. We assume that the Killing vector fields generating stationarity (say $\xi^a$) and axisymmetry (say $\varphi^a$) commute,
\begin{eqnarray}
\nabla_{(a}\xi_{b)} =0= \nabla_{(a}\varphi_{b)} \qquad
[\xi, \varphi]^a =\pounds_{\xi}\varphi^a=0.
\label{f6}
\end{eqnarray}
Let us denote the norms by  $\xi\cdot \xi=-\lambda^2$ and $\varphi\cdot \varphi =+f^2$.

We assume that the 2-dimensional
spacelike surfaces orthogonal to these commuting Killing vector fields form integral submanifolds, which essentially means that the vector fields spanning the subspace form a Lie algebra between themselves, which in turn implies Frobenius-like conditions~\cite{Wald:1984rg},
\begin{eqnarray}
\varphi_{[a}\xi_b\nabla_c \xi_{d]}=\,0\,=\xi_{[a}\varphi_b\nabla_c \varphi_{d]}.
\label{f7}
\end{eqnarray}
Such conditions are purely geometric and are independent of the theory.
The chief difference between the static and stationary axisymmetric spacetime is that, for the later the timelike Killing
vector field is not hypersurface orthogonal, $\xi\cdot\varphi\neq 0$. For convenience, we shall now construct a foliation of the spacetime by a suitable hypersurface orthogonal timelike (non-Killing) vector field. To do this, we define a 1-form $\chi_a$ as
\begin{eqnarray}
\chi_a=\xi_a+\alpha (x)\varphi_a,
\label{f8}
\end{eqnarray}
so that $\chi \cdot \varphi=0$ identically everywhere, giving $\alpha(x) =- (\xi\cdot \varphi)/(\varphi\cdot\varphi)$. The norm of $\chi_a$ is given by
\begin{eqnarray}
\chi_a\chi^a=-\beta^2=-\left(\lambda^2+\alpha^2 f^2\right).
\label{f9}
\end{eqnarray}
This shows that $\chi_a$ is timelike as long as  $\beta^2>0$. However, note that $\chi_a$ is not a Killing vector field in general,
\begin{eqnarray}
\nabla_{(a}\chi_{b)}=\varphi_{(a}\nabla_{b)}\alpha(x).
\label{f10}
\end{eqnarray}
We replace $\xi_a$ in the second of Eq.s~(\ref{f7}) by $\chi_a$ and use it into the first to rewrite both of them as 
$ \varphi_{[a}\chi_b\nabla_c \chi_{d]}=\,0\,=\chi_{[a}\varphi_b\nabla_c \varphi_{d]} $. We can in fact solve for $\nabla_a\chi_b$ and $\nabla_a\varphi_b$ from these two equations, using Eq.s~(\ref{f6}), (\ref{f10}) and $\chi\cdot \varphi=0$ (see~\cite{Bhattacharya:2010vr, Bhattacharya:2011dq, Bhattacharya:2013caa} for details), to get 

\begin{eqnarray}
\nabla_{a}\chi_{b}&=&\beta^{-1}\left(\chi_b\nabla_a\beta-\chi_a\nabla_b\beta\right)+\frac12\left(\varphi_a\nabla_b\alpha+\varphi_b\nabla_a\alpha \right),\nonumber\\
\nabla_{a}\varphi_{b}&=& f^{-1}\left(\varphi_b\nabla_a f-\varphi_a\nabla_bf\right)+\frac{f^2}{2\beta^2}\left(\chi_b\nabla_a\alpha-\chi_a\nabla_b\alpha\right).
\label{f11}
\end{eqnarray}
The first of the above equations shows that $\chi_a$ satisfies the Frobenius condition of hypersurface orthogonality : $\chi_{[a}\nabla_{b}\chi_{c]}=0$. In other words, $\chi^a$ is orthogonal to the family of
3-dimensional spacelike hypersurfaces, say $\Sigma$, containing both $\varphi^a$ and the aforementioned integral 2-submanifolds. 

This particular coordinate independent  $(3+1)$-foliation will be a crucial tool for the rest of our analysis. The metric $g_{ab}$ takes the form in this orthogonal basis, 
\begin{eqnarray}
g_{ab}=-\beta^{-2}\chi_a\chi_b+f^{-2}\varphi_a\varphi_b +\gamma_{ab},
\label{f12}
\end{eqnarray}
where $\gamma_{ab}$ is the metric over the integral submanifolds, orthogonal to both $\chi^a$ and $\varphi^a$. 

Since we have assumed the spacetime to be smooth in our region of interest, all observable quantities (e.g. the Ricci scalar or the trace of the energy momentum tensor) are assumed to be regular as well, which is analogous to the regularity assumption made in the Einstein gravity, e.g.~\cite{nh3}. We assume the matter field to be stationary and axisymmetric as well, i.e. if $X$ is a physical field, we have  $\pounds_{\xi} X=0=\pounds_{\varphi} X$. We assume that all matter fields 
obey the weak and null energy condition, $T_{ab}n^an^b\geq 0$ for any timelike or null vector field.

Note also that the above construction based upon the symmetry holds equally in both $F(R)$ and the Einstein frame, discussed in the preceding subsection. This is just because by our symmetry requirement mentioned above means $ \pounds_{\xi} \phi=0=\pounds_{\varphi} \phi$ and hence the conformal transformation, while acted upon~(\ref{f12}), does not alter the symmetry of the spacetime. Also, it is obvious that such transformation does not alter the integrability condition of the 2-submanifolds as well.

With these, we are now ready to go into studying the horizon properties in stationary axisymmetric spacetimes admitted by a generic $F(R)$ gravity.

\section{Properties of the Killing horizons}
\subsection{The energy condition}
We now proceed to define the Killing horizons. We shall show below that any $\beta^2=0$ compact
hypersurface is a Killing horizon in the sense that $\chi_a$ defined in~(\ref{f8}) becomes Killing there. The method will be similar to the Einstein gravity, i.e. solving the Raychaudhuri equation for the null geodesic congruence on such null surface~\cite{Bhattacharya:2010vr, Bhattacharya:2011dq, Bhattacharya:2013caa}. We first note that as $\beta^2\to 0$, the first of Eq.s~(\ref{f11}) gives,  
\begin{eqnarray}
\chi_{[b}\nabla_{a]}\beta^2\big\vert_{\beta^2\to 0}=\beta^2\partial_{[a}\chi_{b]}\big\vert_{\beta^2\to 0}\to 0,
\label{f13}
\end{eqnarray}
so that on any $\beta^2=0$ hypersurface, $\chi_a$ and $\nabla_a\beta^2$ become parallel,
\begin{eqnarray}
\nabla_a\beta^2=2\kappa(x)\chi_a,
\label{f14}
\end{eqnarray}
where $\kappa(x)$ is a function defined on that hypersurface. Taking the Lie derivative of this equation with respect to $\chi^a$ and using Eq.~(\ref{f6}) and the first of~(\ref{f11}), we find $$\pounds_{\chi}\kappa=0.$$ 
The next step is to construct a congruence of null geodesic over this null surface. Following~\cite{Wald:1984rg}, we define,
 $k_a=e^{-\kappa(x)\tau}\chi_a$, with $\tau$ being the parameter along the null vector field  $\chi^a$ (i.e., $\chi^a\nabla_a\tau:=1$). Thus $k^a$ is null. Using then the first of Eq.s~(\ref{f11}) and~(\ref{f14}), we find $k^a$ satisfies the geodesic equation, $k^a\nabla_a k_b=0$.

The Raychaudhuri equation for the null geodesic congruence reads~\cite{Wald:1984rg}, 
\begin{eqnarray}
\frac{d\theta}{ds}=-\frac12 \theta^2-\sigma_{ab} \sigma^{ab} +\omega_{ab}\omega^{ab}-R_{ab}k^ak^b,
\label{f15}
\end{eqnarray}
where $s$ is an affine parameter, $\theta$, $\sigma_{ab}$ and $\omega_{ab}$ are respectively the expansion, rotation and shear for the congruence, defined on the spacelike compact 2-section of the $\beta^2=0$ hypersurface, orthogonal to $\chi_a$ or $\nabla_a\beta^2$,
\begin{eqnarray}
\theta= \hat{h}^{ab}\widehat{\nabla_ak_b}, \quad \sigma_{ab}= \widehat {\nabla_{(a} k_{b)}}-\frac12 \theta \hat{h}_{ab}, \quad \omega_{ab}=\widehat{\nabla_{[a}k_{b]}},
\label{f16}
\end{eqnarray}
where the `hat' denotes that the quantities are defined on the aforementioned spatial 2-plane and $\hat{h}_{ab}$ is the inverse of the induced 2-metric on that plane. Let us write it as $\hat{h}^{ab}=f^{-2}\varphi^a\varphi^b +\Theta^{-2}\Theta^a\Theta^b$, where $\Theta^a$ is some basis vector orthogonal to the axisymmetric Killing vector. Using Eq.~(\ref{f10}) and the first of Eq.s~(\ref{f11}) we compute
\begin{eqnarray}
k_{[a}\nabla_{b]} k_c= e^{-2\kappa \tau}
\left[\frac12\chi_{(a}\varphi_b \nabla_{c)} \alpha - \chi_c \nabla_a \chi_b - \chi_b \varphi_a\nabla_c \alpha - \chi_b \varphi_c \nabla_a\alpha -\chi_c \chi_{[a} \nabla_{b]} (\kappa \tau )\right].
\label{f17}
\end{eqnarray}
Contracting the above equation respectively with $\hat{h}^{ab}$ and $\varphi^{[a}\Theta^{b]}$ yield $k_a\hat{h}^{bc}\nabla_b k_c=0=k_a \varphi^{[b}\Theta^{c]} \nabla_b k_c$, which, upon comparison with~(\ref{f16}) yields, $\theta=0=\omega_{ab}$.
Likewise, contraction with $\varphi^{(a}\Theta^{b)} $ yields $\sigma_{ab}=\frac12 e^{-\kappa \tau } \varphi_{(a}\widehat{\nabla}_{b)}\alpha$. We plug these results into~(\ref{f15}) to get
\begin{eqnarray}
\frac{f^2}{2}e^{-2\kappa \tau}\left(\widehat{\nabla}_a\alpha\right)\left(\widehat{\nabla}^{a}\alpha\right)+ R_{ab}k^ak^b=0,
\label{f18}
\end{eqnarray}
 which we rewrite using Eq.s~(\ref{f2}), (\ref{f3}) to get
\begin{eqnarray}
\frac{f^2e^{-2\kappa \tau}}{2}\left(\widehat{\nabla}_a\alpha\right)\left(\widehat{\nabla}^{a}\alpha\right)+ \frac{k^ak^b}{F'(R)}\left[8\pi G T_{ab} +\left(\frac{F(R)}{2}+\frac{8\pi G T+2R-RF'(R)}{3}  \right)g_{ab}\right.\nonumber\\\left.+F'''(R)\left(\nabla_a R\right)\left(\nabla_b R\right)+F''(R)\nabla_a\nabla_bR  \right]=0,
\label{f19}
\end{eqnarray}
setting $F(R)=R$ above recovers the result of Einstein's gravity, $R_{ab}=8\pi G\left(T_{ab}-\frac12 T g_{ab}\right) $.
Since $k^a$
is parallel to $\chi^a$, and by our symmetry requirement the Ricci scalar must be stationary and axisymmetric, we have $k^a\nabla_a R=0$. Also, a nonsingular manifold must have non-diverging $R$ and $T$ and non-vanishing $F'(R)$, the inverse of  which plays the role of the modified gravitational coupling in this theory. Putting these all in together, the above equation simplifies
to, 
\begin{eqnarray}
\frac{f^2 e^{-2\kappa \tau}}{2}\left(\widehat{\nabla}_a\alpha\right)\left(\widehat{\nabla}^{a}\alpha\right)+\frac{8\pi G }{F'(R)}T_{ab}k^ak^b
+\frac{F''(R)}{F'(R)}k^ak^b\nabla_a\nabla_b R=0.
\label{f20}
\end{eqnarray}
The first term contains a spacelike inner product, thereby must be a positive definite whereas $T_{ab}k^ak^b\geq 0$ by our choice of the energy condition. The third term seems to have no definite sign and we shall evaluate it explicitly. Since $k^a=e^{-\kappa \tau} \chi^a$, we look at $\chi^a\chi^b\nabla_a\nabla_b R$. We shall compute this term in an infinitesimal neighborhood of the $\beta^2=0$ hypersurface and then will evaluate it on that. We have, since $\chi^a\nabla_a R=0$,
\begin{eqnarray}
\chi^a\chi^b \nabla_a\nabla_b R=-\left(\chi^a\nabla_a\chi^b\right)\left(\nabla_bR\right)=-\left(\nabla_a\beta^2\right)\left(\nabla^aR\right),
\label{f21}
\end{eqnarray}
where we have used Eq.~(\ref{f10}), the orthogonality between $\chi^a$ and $\varphi^a$ and the fact that $\chi^a\nabla_a\alpha=0$ which follows from the commutativity of the two Killing vector fields. 
But on the $\beta^2=0$ surface, $\nabla_a\beta^2$ becomes parallel to $\chi^a$, Eq.~(\ref{f14}). Then it is obvious that 
the above expression vanishes there. Thus we are left only with the first two terms of Eq.~(\ref{f20}), each of which is positive definite. The vanishing sum of them shows that on any compact $\beta^2=0$ hypersurface, we must have
$$T_{ab}\chi^a\chi^b=0~~ {\rm \it and }~~\alpha=~{\rm constant.}$$ The latter means the vector field $\chi^a=\xi^a+\alpha \varphi^a$ is a null Killing vector field there and hence all such surfaces are Killing horizons of this theory. The emphasis on the compactness is due to the fact that we have taken the axisymmetric Killing vector field  $\varphi^a$ to be one of the spatial generators of that hypersurface. These conditions are similar to that of the Einstein's gravity~\cite{Bhattacharya:2010vr, Bhattacharya:2011dq, Bhattacharya:2013caa}. Our analysis thus shows the universality  of the Killing horizons of these two theories.

Having seen that $\alpha$ is a constant tangent to the Killing horizon, let us now see how it behaves off the horizon, which will be useful for our later purpose. In order to see this, let us choose $\nabla_a\beta^2$ to be one of the basis vector fields. This is linearly independent of $\chi^a$ and $\varphi^a$ by virtue of the commutativity of the Killing vector fields, $\chi^a\nabla_a\beta^2=0=\varphi^a\nabla_a\beta^2$. Then, since we have assumed the spacetime to be smooth, Eq.~(\ref{f14}) shows that $\nabla_a\beta^2$ becomes null as ${\cal O}(\beta^2)$ in the infinitesimal neighborhood of the horizon and
as well,
$$\kappa^2(x)= \lim_{\beta^2\to 0}\frac{\left(\nabla_a\beta^2\right)\left(\nabla^a\beta^2\right)} {4\beta^2}.$$
Let us denote $\nabla_a\beta^2 $ by $Z_a$ and let $Z$ be the local parameter along it, such that $Z^a\nabla_aZ:=1$.
Then we can replace the numerator of the above expression by $d\beta^2/dZ$, which must be at least ${\cal O}(\beta^2)$ in order to make $\kappa$ finite ($\kappa \neq 0$ only when the numerator is ${\cal O}(\beta^2)$).  Now, if we evaluate $\left(\nabla_a\alpha\right)\left(\nabla^a\alpha\right)$, the part tangent to the horizon vanish as earlier, whereas the part along $Z^a$
gives $\left(\frac{d\alpha}{dZ}\right)^2$ divided by the norm of $Z^a$, which, as we have seen, vanish at least as ${\cal O} (\beta^2)$. We further have $\frac{d\alpha}{d Z}=\frac{d\alpha}{d\beta^2}\frac{d\beta^2}{dZ}$. Since $\alpha=-\frac{\xi\cdot \varphi}{\varphi\cdot \varphi}$, it is reasonable to assume that it is analytic in $\beta^2$, which ensures the finiteness 
of the term $\frac{d\alpha}{d\beta^2}$\footnote{In other words, if this is not the case, the angular velocity on the horizon could be infinite, due to the existence of of terms of negative powers of $\beta^2$.}. Then it is clear that $ \frac{1}{Z\cdot Z}\left(\frac{d\alpha}{dZ}\right)^2$ vanishes on the horizon as $Z\cdot Z$, which at least ${\cal O}(\beta^2)$, where the equality holds for $\kappa \neq 0$. This means that 
$$\left(\nabla_a\alpha\right)\left(\nabla^a\alpha\right)\vert_{\beta^2=0}=0$$


To summarize, for a generic $F(R)$ gravity, we have found for a smooth stationary axisymmetric spacetimes a coordinate independent $(3+1)$-foliation of the spacetime. The timelike foliation vector field $\chi^a$ becomes Killing whenever it becomes null, thereby
giving the Killing horizons of the theory. Clearly, apart from the black hole, if there is a cosmological event horizon as well, it will be defined in the same footing via the vector field $\chi^a$.

\subsection{The constancy of $\kappa$} 
Using the above result, like the General Relativity~\cite{Wald:1984rg}, it is now easy to prove the constancy of the function $\kappa(x)$ (Eq.~(\ref{f14})), the so called surface gravity, on any Killing horizon irrespective of whether the effective gravitational coupling $\sim (F'(R))^{-1}$ is a constant there or not. From now on, we shall assume that $\kappa \neq 0$.  

Since $\chi_a$ is Killing on the horizon, $\nabla_{(a}\chi_{b)}\vert_{\beta^2=0}=0$,  and is hypersurface orthogonal everywhere (cf., the first of~(\ref{f11})), we may rewrite on the horizon the Frobenius condition as,
\begin{eqnarray}
\chi_{[a}\nabla_b\chi_{c]}=\chi_a\nabla_b\chi_c+ \chi_b\nabla_c\chi_a+\chi_c\nabla_a\chi_b=0.
\label{f22}
\end{eqnarray}
Since the Killing vector field  $\chi^a$ (or $\nabla_a\beta^2$) is normal to the horizon, the relevant derivative operator
tangent to it would be $\chi_{[a}\nabla_{b]}$~\cite{Wald:1984rg}. Then our precise goal would be to prove that $\chi_{[a}\nabla_{b]}\kappa=0$.
Using $\beta^2=-\chi^a\chi_a$ in Eq.~(\ref{f14}), acting $\chi_{[a}\nabla_{b]}$ on it and using the Killing identity $\nabla_a\nabla_b \chi_c=-R_{bca}{}^{d}\chi_d$,  we have
\begin{eqnarray}
\chi_c \chi_{[a}\nabla_{b]}\kappa+\kappa \chi_{[a}\nabla_{b]}\chi_c=\left(\chi_{[a}\nabla_{b]}\chi^d\right)\left(\nabla_d\chi_c\right)+\chi^d R_{dc[a}{}^{e}\chi_{b]}\chi_e.
\label{f23}
\end{eqnarray}
Eq.~(\ref{f22}) shows  by virtue of the Killing equation on the horizon, $\chi_{[a}\nabla_{b]} \chi_c= \chi_c\nabla_b\chi_a$. Using this along with~(\ref{f14}) into the above equation it is easy to see that
the second term on the left hand side exactly equals the first term on the right hand side. We next substitute for the decomposition of the Riemann tensor,
\begin{eqnarray}
R_{abcd}=C_{abcd}+\left(g_{a[c}\,R_{d]b}-g_{b[c}\,R_{d]a} \right) -\frac{R}{3}\left(g_{ac}\,g_{bd}-g_{bc}\,g_{ad} \right),
\label{f24}
\end{eqnarray}
into~(\ref{f23}), the Ricci scalar terms go away to yield
\begin{eqnarray}
\chi_c \chi_{[a}\nabla_{b]}\kappa = \chi^d\chi_e C_{dc[a}{}^{e}\chi_{b]}+ \chi_c\chi^dR_{d[a}\chi_{b]}-R_{ed}\chi^e\chi^dg_{c[a}\chi_{b]}.
\label{f25}
\end{eqnarray}
But the discussions of the preceding subsection has shown, on the horizon\\ $R_{ab}\chi^a\chi^b \equiv (8\pi G/F'(R))\, T_{ab}\chi^a\chi^b=0$. Let us now focus on the conformal tensor term. For the null geodesic congruence $\{k^a\}$, we have~\cite{Wald:1984rg},
\begin{eqnarray}
k^c\nabla_c \sigma_{ab}=-\theta \sigma_{ab}+\widehat{C_{cbad}k^ck^d},
\label{f26}
\end{eqnarray}
where the `hat', as earlier denotes that the components have been evaluated (including $\theta$ and $\sigma_{ab}$ themselves) on the spatial 2-section of the horizon. We already have proven that $\theta=0=\sigma_{ab}$. Now we shall prove
the left hand side of the above equation is vanishing, too. Since $k^a=e^{-\kappa(x)\tau}\chi^a $, let us evaluate $\chi^c\nabla_c \sigma_{ab}$, which equals, using our previous results, 
$$\frac12\chi^c\nabla_c \left(e^{-\kappa(x)\tau }\varphi_{(a}\widehat{\nabla}_{b)}\alpha \right)$$
Recalling $\chi^a\nabla_a\tau:=1$ and $\pounds_{\chi}\kappa=0=\pounds_{\chi}\alpha$, the above expression reduces to 
$$\frac12\left[-\sigma_{ab}+ e^{-\kappa(x)\tau} \left(\left(\chi^c\nabla_c\varphi_{(a}\right)\widehat{\nabla}_{b)}\alpha- \varphi_{(a}\left(\widehat{\nabla}_{b)} \chi^c\right)\nabla_c \alpha \right)\right]$$
We substitute Eq.s~(\ref{f11}) into the above and `hat' both the indices $a$ and $b$. The commutativity of the Killing vector fields gives $\pounds_{\chi}\alpha=0=\pounds_{\varphi}\alpha$. Putting these all in together and using $\sigma_{ab}=0 $, the above expression reduces to, 
\begin{eqnarray}
\frac12e^{-\kappa(x)\tau}\left[f^2\left(\widehat{\nabla}_a\alpha\right)\left(\widehat{\nabla}_b\alpha\right)+2\varphi_a\varphi_b \left(\nabla_c\alpha\right)\left(\nabla^c\alpha\right) \right],
\label{f27}
\end{eqnarray}
note that since $\varphi_a$ is already tangent to the spatial 2-section of the horizon, we did not need to `hat' it.
From the discussions of the preceding subsection, it is now clear that the above expression vanishes, leaving us only with 
the second term on the right hand side of Eq.~(\ref{f25}). Substituting for $R_{ab}$ from Eq.s~(\ref{f2}), (\ref{f3}) into this we get
\begin{eqnarray}
\chi_{[a}\nabla_{b]}\kappa = \frac{\chi^d} {F'(R))}\left[8\pi G\, T_{d[a} +\left(\frac{F(R)}{2}+\frac{8\pi G T+2R-RF'(R)}{3}  \right)g_{d[a}\right.\nonumber\\\left.+F'''(R)\left(\nabla_d R\right)\left(\nabla_{[a} R\right)+F''(R)\nabla_d\nabla_{[a}R  \right] \chi_{b]}.
\label{f28}
\end{eqnarray}
All terms except the one containing $T_{ab}$ vanishes, as earlier to get $\chi_{[a}\nabla_{b]}\kappa =\frac{8\pi G }{ F'(R)} \chi^dT_{d[a}\chi_{b]}$. The fact derived in the earlier subsection, $T_{ab}\chi^a\chi^b=0$ does not automatically guarantee
that $T_{ab}\chi^b$ is parallel to $\chi_a$. However, since $\chi_a$ is timelike and hypersurface orthogonal everywhere and is Killing and null on the horizon, the horizon and its infinitesimal neighborhood's geometry is similar to that of the static. In that case the time reversal invariance will rule out any cross term in $T_{ab}\chi^b$. This guarantees that the right hand side of~(\ref{f28}) vanishes thereby proving that $\kappa$ is a constant over the horizon.

Thus we have shown that the horizon or the near horizon geometry for the General Relativity and $F(R)$ theories are formally identical.

We shall end this section with a comment on the entropy of stationary black holes in this theory and its comparison with the corresponding Einstein frame~\cite{Faraoni, Chatterjee:2010gp}. We can use the analogue of the Gibbons-Hawking-York surface counterterm, $-\frac{1}{8\pi G}\int_{\rm Boundary} F'(R) K$, where $K$ is the extrinsic curvature of the boundary, which is the horizon in this case. Using the variation of this boundary term under the action of various diffeomorphism generating vector fields that retain the near horizon structure, one can compute the entropy, cf. the formalism developed in~\cite{Majhi1, Majhi2, Majhi3}.  The entropy turns out to be $S=\frac{1}{4\pi G}\int F'(R) d\Sigma^{(2)}$, where the integration measure is over the spatial compact section of the horizon. 
  
For a general stationary axisymmetric spacetime, as we have emphasized, there is no reason for which $F'(R)$ could be a constant on the horizon. This shows that the entropy in the $F(R)$ theory may not scale like the horizon's area. The only way to relate this to the area seems to define an averaging over the horizon,  $\langle F'(R)\rangle=(\int F'(R) d\Sigma^{(2)})/(\int d\Sigma^{(2)}) $. With this seemingly {\it ad hoc} prescription,  the entropy becomes
$S=\frac{A}{4\pi G_{\rm eff.}}$, where $A$ is the horizon area and $G_{\rm eff. }= G/\langle F'(R)\rangle$.

Nevertheless, it is easy to see that the entropy of black holes would be the same in both $F(R)$ and the Einstein frames, at least numerically, if not functionally. This follows from the surface counterterm for~(\ref{f5}), which is just  $-\frac{1}{8\pi G}\int_{\rm Boundary} \widetilde{ K}$. This gives the entropy to be~\cite{Majhi1, Majhi2, Majhi3}, $ S=\frac{1}{4\pi G}\int d\widetilde{\Sigma}^{(2)}$. But from the conformal transformation it is clear that  $d\widetilde{\Sigma}^{(2)}=F'(R)d\Sigma^{(2)}$, thereby proving the equality. Nevertheless, we must emphasize here that the aforementioned difference  between the functional behaviour of the horizon entropy  is perhaps the most prominent qualitative distinction between frames~(\ref{f1}) and~({\ref{f5}}).

However, while going from the Jordan Brans-Dicke~(\ref{f4}) to Einstein's frame~(\ref{f5}) one throws away a total divergence of the scalar field, $\Box \Phi $, e.g.~\cite{Canate:2015dda}. Such terms always arise when we consider a conformal transformation $\widetilde{g}_{ab}=\Omega^2 g_{ab}$ and compute the Ricci tensor $\widetilde{R}$ in terms of $R$ and $\Omega$~\cite{Wald:1984rg}. The crucial point is, such term might also lead to boundary terms, determined by the normal derivative of the scalar field on the horizon which will be relevant for the scalar hairy black holes. In the presence of such terms, it is not obvious as above that the two entropies should be the same. We wish to return to this issue in detail in a future work.    

Having discussed generic local properties of Killing horizons in $F(R)$ gravity, we shall now move onto the global no hair theorems.

\section{The no hair theorems}
\subsection{Scalar field}
The discussions on the no hair theorems below would chiefly be based on the techniques of~\cite{Bhattacharya:2011dq} developed for the General Relativity, to which we shall often refer to the reader for further details.   Let us start by considering a real scalar field  $\Psi$ moving in a potential $V(\Psi)$. 
\begin{eqnarray}
\Box \Psi-V'(\Psi)=0,
\label{f29}
\end{eqnarray}
where a `prime' denotes differentiation with respect to $\Psi$ once. We shall project this equation onto the spacelike hypersurface $\Sigma$, orthogonal to $\chi^a$. The projector which projects spacetime tensors onto $\Sigma$ is given by
$$h_a{}^b=\delta_a{}^b+\beta^{-2}\chi_a\chi^b.$$ We write the inverse metric as $g^{ab}=h^{ab}+\beta^{-2}\chi^a\chi^b$.
Then since by our symmetry assumption the scalar field is stationary and axisymmetric, we must have $\chi^a\nabla_a\Psi=0$ (see~\cite{Smolic:2015txa}, for a further formal discussion on the validity of such symmetry requirement).
Then it turns out that $\Box \Psi= \frac{1}{\beta \sqrt{h}}\partial_a\left(\beta \sqrt{h} h^{ab}\partial_b \Psi\right) $, where $h$ is determinant of the induced metric $h_{ab}$ on $\Sigma$ (this equals $f^{-2}\varphi_a\varphi_b+\gamma_{ab}$ in~(\ref{f12})). 
If $D_a$ is the covariant derivative operator on $\Sigma$ associated with the induced metric $h_{ab}$, then
$D_aD^a \Psi=\frac{1}{\sqrt{h}}\partial_a\left(\sqrt{h}h^{ab}\partial_b \Psi \right)$. Comparing this with $\Box \Psi$,
we have $$\Box \Psi=\frac{1}{\beta} D_a\left(\beta D^a \Psi \right). $$ We substitute this into~(\ref{f29}) and multiply with $V'(\Psi)$ and then integrate by parts to find 
\begin{eqnarray}
\int_{\partial \Omega}\beta V' (\Psi)n^aD_a \Psi -\int_{\Sigma}\beta\left[V''(\Psi) \left(D_a\Psi\right)\left( D^a\Psi\right)+ V'^2(\Psi)\right]=0,
\label{f30}
\end{eqnarray}
where the surface integral are respectively taken on the horizon $(\beta=0)$ and infinity, both of which vanish. The inner product $\left(D_a\Psi\right)\left( D^a\Psi\right)$ is spacelike and hence positive definite. 
Thus if the potential is convex $V''(\Psi)\geq 0$, it turns out from the volume integral of~(\ref{f30}) that $\Psi$
is a constant sitting on the minimum of the potential. This is the usual no hair result. Thus real scalars with a convex potential will definitely satisfy this theorem in generic  $F(R)$ gravity. However, we shall see below that this will not be the case for the  massive vector field for rotating black holes in this theory.

\subsection{The massive vector field}
The Proca massive vector field has Lagrangian density, $${\cal L}=-\frac14 F_{ab}F^{ab}-\frac12 m^2A_bA^b,$$ where $F_{ab}=\nabla_{[a}A_{b]}$. The equation of motion of reads $$\nabla_aF^{ab}-m^2A^b=0.$$ To deal with this theory in stationary axisymmetric spacetimes, we need,
in addition to the projector $h_a{}^b$ mentioned above, the projector $\gamma_a{}^b$ which projects tensors onto the spacelike 
integral submanifolds orthogonal to both $\chi^a$ and $\varphi^a$  mentioned in Sec.~2,$$\gamma_a{}^b=\delta_a{}^b+\beta^{-2}\chi_a\chi^b-f^{-2}\varphi_a\varphi^b. $$
The prescription is now the following~\cite{Bhattacharya:2011dq}. We first project the equation of motion onto the family of spacelike hypersurfaces, $\Sigma$, to get  
\begin{eqnarray}
D_a\left(\beta f^{ab}\right)=m^2\beta a^b+\frac12 \varphi^be^c\nabla_c\alpha,
\label{f31}
\end{eqnarray}
where the projections are : $f_{ab}=h_a{}^ch_b{}^dF_{cd}$ and $a_b=h_b{}^cA_c$ and $e^a=\beta^{-1}\chi_b F^{ab}$. The last term on the right hand side comes from the Lie derivative of $A_b$. Since the vector field is not Killing except on the horizon, it survives.

The goal of the no hair proofs are to construct suitable positive definite vanishing integrals, to show that the fields vanish. However, we cannot possibly do this for~(\ref{f31}), due to the existence of the last term on the right hand side, which does not have any definite sign. Thus we further project that equation onto the integral 2-submanifolds, using the projector $\gamma_a{}^b$. Since these submanifolds are orthogonal to both $\chi^a$ and $\varphi^a$, the last term of Eq.~(\ref{f31}) goes away by this operation,
giving  $$\overline{D}_a\left(f\beta \overline{f}^{ab}\right)=m^2f\beta  \overline{a}^b, $$
where $\overline{a}_b=\gamma_b{}^ca_c$ and $\overline{f}_{ab}=\gamma_a{}^c\gamma_b{}^df_{cd}$ and $\overline{D}$ is the induced derivative on the 2-submanifolds. We now contract the above equation with $\overline{a}_b$ and integrate by parts between the horizon and infinity to obtain $\overline{a}_b=0$ over the integral submanifolds orthogonal to both $\chi^a$ and $\varphi^a$. 

Thus we are left with only two components of $A_a$, directing along $\chi_a$ and $\varphi_a$. We write $A_a=\Psi_1\chi_a+\Psi_2\varphi_a$. We substitute this ansatz into the Lagrangian and use~(\ref{f11}). The resulting two equations of motion
corresponding to $\Psi_1$ and $\Psi_2$, when integrated by parts between the horizon and infinity as earlier and added together, yield 
\begin{eqnarray}
\int_{\Sigma}\beta\left[\left(\beta D_a\Psi_1+2\Psi_1D_a\beta\right)^2+\left(f D_a\Psi_2+2\Psi_2D_af\right)^2 -\frac{f^4 \Psi_2^2}{\beta^2}\left(D_a\alpha\right)\left(D^a\alpha\right)+m^2\left(\beta^2\Psi_1^2+f^2\Psi_2^2\right)\right]=0,\nonumber\\
\label{f32}
\end{eqnarray}
all of the integrands, except the third one is positive definite. The third term should intuitively be interpreted as the centrifugal effect due to spacetime rotation. In order to estimate this term, we consider the Killing identity, $\Box \varphi_a=-R_a{}^b\varphi_b$. We contract this with $\varphi^a$ and use the second of~(\ref{f11}), multiply the resulting equation with $\Psi_2^2$ and integrate by parts as earlier to get     
$$\int_{\Sigma} \beta \left[4f\Psi_2\left(D_a\Psi_2\right) \left(D^af\right)+4\Psi_2^2  \left(D^af\right) \left(D_af\right)  -\frac{f^4 \Psi_2^2}{\beta^2}\left(D_a\alpha\right)\left(D^a\alpha\right)-2\Psi_2^2 R_{ab}\varphi^a\varphi^b\right]=0$$
Subtracting the above from Eq.~(\ref{f32}) we get
\begin{eqnarray}
\int_{\Sigma}\beta\left[\left(\beta D_a\Psi_1+2\Psi_1D_a\beta\right)^2+f^2\left(D_a\Psi_2\right) \left(D^a\Psi_2\right)+ 2\Psi_2^2 R_{ab}\varphi^a\varphi^b+m^2\left(\beta^2\Psi_1^2+f^2\Psi_2^2\right)\right]=0.
\label{f33}
\end{eqnarray}
Clearly, the validity of the no hair theorem now solely depends upon the positivity of $R_{ab}\varphi^a\varphi^b$.

It is the point where the thing deviates from the General Relativity. For the latter, we have $R_{ab}=8\pi G \left(T_{ab}-\frac12T g_{ab}\right)$. For the Proca energy-momentum tensor we have always, $R_{ab}\varphi^a\varphi^b\geq 0$~\cite{Bhattacharya:2011dq}. However for the $F(R)$ theory, Eq.s~(\ref{f2}), (\ref{f3}) show that, at least apparently, we cannot get any such definite result.

In order to reach some physically reasonable conclusion, at this point we use the correspondence between the $F(R)$ and the Einstein frames, which seems to require, $F''(R)>0$~\cite{Canate:2015dda}. Let us now see how Eq.~(\ref{f33}) will look like when written in that frame. For the term $\int d^4x \sqrt{-g} \left(-\frac12 m^2 g^{ab}A_{a}A_b\right) $, the mass  will be replaced with 
$m^2\to m^2/F'(R)$ (cf., the  discussions of 2.1). We denote this as $m^2(x)$. Everything else would formally be the same, as we assumed that the conformal transformation is such that the isometries, the Killing horizons and the asymptotic structures remain the same in both the frames.

Now, having written everything in the Einstein frame, it is obvious that  $R_{ab}\varphi^a\varphi^b\big\vert_{\rm Einstein}$ in Eq.~(\ref{f33}) will have two contributions -- one from the Proca field and the other from the scalar field $\Phi$ generated in the Einstein's frame~(\ref{f5}). It turns out using the axisymmetry of the scalar field $\Phi$ that,
$$R_{ab}(\Phi)\varphi^a\varphi^b=\left(T_{ab}(\Phi)-\frac12 T(\Phi)\widetilde{g}_{ab}\right)\varphi^a\varphi^b=\widetilde{f}^2 U(\Phi),$$ where we have put `tilde' since we have written the metric in the Einstein frame as $\widetilde{g}_{ab}$.

This, along with the fact that $R_{ab}\varphi^a\varphi^b$ is always positive definite in the Einstein frame for the Proca field leads to the conclusion that if $U(\Phi)\geq 0$, Eq.~(\ref{f33}) will force the Proca no hair theorem to hold. If the field vanishes in the Einstein frame, it would vanish also in the $F(R)$ frame. 

\subsection{A simple corollary}
Before we end, we shall present a simple corollary  for the Starobinsky model~\cite{Starobinsky} : $F(R)=R+ \gamma_1 R^2$, where the parameter $\gamma_1 >0 $.  Eq.~(\ref{f3}), with conformally invariant matter field ($T=0$) becomes,
\begin{eqnarray}
\Box R= \frac{R}{\gamma_1}.
\label{f34}
\end{eqnarray}
Following the earlier procedure for the scalar field, we project the above equation onto the spatial hypersurface $\Sigma$, multiply with $R$, (using, by the stationarity and axisymmetry, $\chi^a\nabla_a R=0$) and integrate by parts to get
\begin{eqnarray}
\int_{\partial \Omega}\beta  R n^a \nabla_a R=\int _{\Sigma} \beta \left[(D_a R)(D^a R)+ \frac{R^2}{\gamma_1}  \right],
\label{f35}
\end{eqnarray}
where, as earlier the boundary integral is done from the horizon $(\beta =0)$ up to the the asymptotically flat region. Since, $R\sim {\cal O}(\frac{1}{r^3})$ as $r\to \infty$, the boundary integral vanish.
Then, since $\gamma_1$ is positive, the above equation shows that the Ricci scalar is not only a constant, but also it vanishes  {\it everywhere} in our region of interest. We may plug this result back into Eq.~(\ref{f2}), to get only the Einstein equations. Thus we may conclude that for the Starobinsky model, the only asymptotically flat, electrovacuum and stationary axisymmetric 
spacetime is the Kerr-Newman family.

\section{Summary and outlook}
\noindent
In this work we have investigated some basic properties of rotating Killing horizons for generic  $F(R)$ gravity
under some suitable geometric framework. We have proved similarities of those Killing horizons with that of Einstein's gravity, including in particular, the constancy of the surface gravity. This analysis thus shows the universality of the formal horizon properties of the two theories. Using this framework, we have discussed the basic no hair theorems in such theories. For the Proca field in particular, assuming the correspondence between frames~(\ref{f1}) and~(\ref{f5}), we have shown that the no hair theorem holds if the potential of the scalar field generated in the Einstein's frame is a positive definite quantity.

Investigation of Killing horizons is an integral and essential part in the study of black hole physics. While the primary motivation behind the study of alternatives to $\Lambda{\rm CDM}$ is to understand Dark Energy/Dark Matter, any such theory should be checked against the stationary black hole solutions and their uniqueness properties as well. This gives us an interesting physical arena to realize how Einstein's theory is qualitatively/quantitatively different (or, similar) from those viable alternatives (e.g.~\cite{Vigeland} and Ref.s therein). The $F(R)$ gravity could be physically relevant in particular, via some hitherto unknown mechanism of quantum gravity, in the context of very small black holes.  Note also that since the imaginary part the quasinormal modes for black holes in Einstein's theory could be given by the surface gravity $\kappa$ of the event horizon (e.g.~\cite{Konoplya} and Ref.s therein), the constancy of  $\kappa$ in the present case could be an indication that generic $F(R)$-black holes are also endowed with quasinormal mode spectra qualitatively similar to that of Einstein's.

We note here an interesting thing in the context of the no hair theorem -- for static and spherically symmetric spacetimes, we can just put $\Psi_2=0$ in Eq.~(\ref{f32}), because in that case the only relevant component 
for the vector field is $A_t\equiv \Psi_1$. In that case the no hair theorem holds without any further condition, unlike the stationary axisymmetric spacetimes. This could be a possible qualitative
difference between rotating and non-rotating spacetimes in $F(R)$ theory, not present in the Einstein gravity. Nevertheless, it might also be possible that such theorem would also hold for the rotating $F(R)$ spacetime as well without any restriction on $U(\Phi)$, due to some additional physical conditions or identities involving $F(R)$ and its derivatives. However, so far it remains elusive to us. In any case, it is evident that $U(\Phi)\geq 0$ is not a very strong condition -- after all, it is just necessary to ensure the stability of the scalar field in the Einstein frame. In other words, the current work rules out any hairy solution for a real massive vector field for all $U(\Phi)\geq 0$, which seems to be important in its own right.

The next interesting thing would be to investigate the area theorems in such theories, as far as the horizon properties are concerned. We hope to address this issue sometimes in the future.

\section*{Acknowledgement}
The author sincerely acknowledges T.~Padmanabhan for suggesting various things on this work, for exciting discussions, for reading the manuscript and encouragement. He thanks Kinjalk Lochan for useful
 discussions and for a careful critical reading of the manuscript. He also acknowledges anonymous referee for various useful comments and suggestions.


\end{document}